# Non-polar (11-20) InGaN quantum dots with short exciton lifetimes grown by metal-organic vapor phase epitaxy

*Tongtong Zhu\*,1,a, Fabrice Oehler\*a, Benjamin P. L. Reidb, Robert M. Emerya, Robert A. Taylorb, Menno J. Kappersa, and Rachel A. Olivera*

[a]*Department of Materials Science and Metallurgy, University of Cambridge, Pembroke Street, Cambridge CB2 3QZ, United Kingdom*
[b]*Department of Physics, University of Oxford, Parks Road, Oxford, OX1 3PU, United Kingdom*

**ABSTRACT**

We report on the optical characterization of non-polar $a$-plane InGaN quantum dots (QDs) grown by metal-organic vapor phase epitaxy using a short nitrogen anneal treatment at the growth temperature. Spatial and spectral mapping of sub-surface QDs have been achieved by cathodoluminescence at 8 K. Microphotoluminescence studies of the QDs reveal resolution limited sharp peaks with typical linewidth of 1 meV at 4.2 K. Time-resolved photoluminescence studies suggest the excitons in these QDs have a typical lifetime of 538 ps, much shorter than that of the $c$-plane QDs, which is strong evidence of the significant suppression of the internal electric fields.

*\*T. Zhu and F. Oehler contributed equally to this work*
[1]*To whom correspondence should be addressed tz234@cam.ac.uk*

Nitride-based quantum dots (QDs) have been suggested to be very promising candidates for a variety of optoelectronic applications and for the investigation of light-matter interactions[1,2]. The large band offset between QD and matrix material and large exciton binding energy offer the potential of single photon emission up to room temperature or at least up to temperatures easily accessible by thermoelectric cooling[3]. InGaN QDs are particularly attractive since they provide a convenient route to emission in the blue and green spectral region by varying the alloy content, an emission range well-matched to commercially available ultrafast single photon detectors.

The wurtzite crystal structure of nitride materials leads to very large piezoelectric constants and hence for *c*-plane structures significant electric fields across strained QDs. This allows tuning of the QD emission across a wide range of wavelengths and control of the oscillator strength by application of an external bias[4]. The disadvantage of the internal electric fields, however, is the spatial separation of the electron and hole wavefunctions that reduces the radiative recombination efficiency due to the quantum confined Stark effect (QCSE), and which leads to long exciton lifetimes increasing the time-jitter on the emission from a single photon source and reducing the available repetition rate. These in-built electric fields may also be responsible for spectral diffusion (the variation of the QD emission wavelength on a timescale of a few seconds), which makes maintaining resonance between a QD and an optical cavity difficult[5]. Growth of InGaN QDs along non- and semi-polar orientations is thus of interest since the internal electric fields can be greatly reduced or eliminated[6,7]. Although there are a few reports on the growth of InGaN nanostructures on non-polar *m*-plane (1-100) by metal-organic vapor phase epitaxy (MOVPE)[8] and (11-22) semi-polar orientation by molecular beam epitaxy[9], little progress has been reported on the growth of non-polar *a*-plane (11-20) InGaN QDs.

In this letter, we demonstrate the growth of non-polar *a*-plane (11-20) InGaN QDs by MOVPE using a short nitrogen anneal treatment at the InGaN growth temperature to form nanostructures. Low-temperature cathodoluminescence (CL) and micro-photoluminescence (µPL) have spatially and spectrally resolved these *a*-plane QDs. Time-resolved single photon counting suggests much shorter lifetimes for excitons confined in *a*-plane InGaN QDs compared to those of the *c*-plane QDs.

The samples were grown by MOVPE in a 6 x 2 in. Thomas Swan close-coupled showerhead reactor on *r*-plane (1-102) sapphire substrates. Trimethylgallium, trimethylindium, and ammonia were used as precursors. To improve the surface morphology and material quality, *a*-plane GaN pseudo-substrates were prepared using the epitaxial lateral overgrowth (ELOG) technique. The seed layer consists of 1 µm of GaN grown with a V/III ratio of ~60 at 1050 °C, following a 30 nm GaN nucleation layer grown at 540 °C and 500 Torr. A $SiO_2$ layer of 100 nm thickness was deposited onto the seed layer *ex-situ* by e-beam evaporation. Conventional wet chemistry lithography was then used to pattern the $SiO_2$ layer into 5 µm-wide stripes and 5 µm-wide window openings parallel to the [1-100] GaN direction. Regrowth was initially carried out with a V/III ratio of 60 at 1050 °C and 100 Torr to promote lateral growth and completed by the growth of a further 2 µm of GaN with a V/III ratio of 740 to improve luminescence properties[10]. This type of as-grown ELOG GaN pseudo-substrate was characterized previously by transmission electron microscopy[11]. In summary, in the window region, the dislocation density was $(1.0\pm0.06) \times 10^{10}$ $cm^{-2}$ and the density of basal plane stacking faults (BSFs) was $(2.6\pm0.3) \times 10^5$ $cm^{-1}$. The defect density in the wing regions is much lower. The +*c* wing has a very low dislocation density (less than $1 \times 10^6$ $cm^{-2}$), and a reduced density of BSFs ($(2.0\pm0.7) \times 10^4$ $cm^{-1}$), with the majority of the observed BSFs being clustered at the boundary between the wing and window regions, and an almost BSF-free region being found closer to the coalescence boundary. The -*c* wing is narrower and defects are distributed across it more uniformly giving a BSF density in this wing of $(2.4\pm0.5) \times 10^5$ $cm^{-1}$ and a dislocation density of $(9\pm1) \times 10^8$ $cm^{-2}$.

InGaN epilayers of ~10 ML thickness were grown at 695 °C with $N_2$ as the carrier gas at a reactor pressure of 300 Torr. Immediately after the InGaN epilayer growth, samples were annealed for 30 s in an $N_2$ atmosphere under the same conditions of temperature and pressure as the InGaN epilayer growth. For QD formation, the annealed epilayer was exposed to ammonia and capped with ~10 nm of GaN at 695 °C in $N_2$ and another ~10 nm of GaN at 1050 °C in $H_2$.

Samples for atomic force microscopy (AFM) studies were cooled to room temperature immediately after the $N_2$ anneal step. AFM was carried out using a Dimension 3100 instrument in tapping mode using RTESP tips from Bruker Nano. CL studies were performed on a liquid helium cooled stage at 8 K in a Philips XL30s scanning electron microscope operating at 5 kV and equipped with a Gatan MonoCL4 system. Micro-PL measurements were carried out using a two-photon excitation technique[12] employing a picosecond mode-locked Ti-sapphire laser emitting at 790 nm. Samples were mounted in a cold-finger cryostat that could be cooled down to 4.2 K and the laser was focused through a microscope objective lens to a spot size of ~1 µm with an excitation power density of 7.64 MW cm$^{-2}$. Luminescence from the samples was dispersed using a 1200 l/mm grating in a 0.30 m spectrometer with a Peltier cooled charge-coupled detector (CCD), which gives a spectral resolution of ~800 µeV. Time-resolved PL measurements were also taken at 4.2 K using the same monochromator and employing a time-correlated single photon counting system based on a fast photomultiplier tube, giving a time resolution of 130 ps.

Figure 1(a) shows an AFM image of an uncapped and annealed InGaN epilayer (in $N_2$ for 30 s at 695 °C) with part of the ELOG +*c*-wing and window areas visible. Small three-dimensional nanostructures have been formed on the surface of both the highly defective window and low defect density wing regions with an average density of ~3 x 10$^8$ cm$^{-2}$ and an average height of 7±3 nm. The detailed AFM scan in Fig. 1(b) shows the interface between a window and wing region. In the window region, triangular shaped pits, which correspond to partial dislocations threading through the film[13] are observed. It appears that the size and distribution of the nanostructures is fairly similar across the window and wing regions, suggesting that the surface morphology of the underlaying *a*-plane GaN is not playing an important role in the nanostructure formation. The AFM data in Fig. 1(b) also show that the remaining epilayer consists of a network of interlinking strips (~10 nm in width) aligned approximately parallel to the [0001] direction. Room temperature CL studies of the annealed epilayer reveal a broad emission band between 410 and 450 nm originating from the strip network, indicating that the strips contain InGaN. An interlinking network of InGaN strips is also seen in addition to the small nanostructures when *c*-plane InGaN epilayers are annealed in nitrogen, but in that case the network is coarser and more disordered[14].

We have investigated the effects of etching these nanostructures in $HCl/H_2O$ (1:3) and a significant reduction in the density of the nanostructures has been observed by AFM. Therefore, we suggest that the nanostructures seen in Fig. 1(a) and (b) are metallic Indium/Gallium droplets formed by the decomposition of the InGaN epilayer, which has been enhanced by the absence of active nitrogen from ammonia gas during the annealing process. The formation mechanism of the nanostructures appears to be analogous to the modified droplet epitaxy that has been used to obtain *c*-plane InGaN QDs[14].

Figure 2(a) shows a panchromatic CL image of an annealed and capped *a*-plane annealed InGaN layer taken at 8 K. Bright spots are observed in both window and wing regions with an average density of ~1 x $10^8$ $cm^{-2}$, i.e., a similar nanostructure density to that measured by AFM. However, it is worth noting that some bright spots tend to have a tail towards the [000-1] direction (indicated by the red arrows). Further microstructural studies of this phenomenon are currently underway. In addition, there are linear emission features along the [1-100] direction present in the highly defective window region, which are associated with the luminescence from basal plane stacking faults[10].

We investigated the luminescence properties of individual bright spots by taking a spatial CL intensity map of an area consisting of 33 x 22 pixels with a step size of 180 nm [Fig. 2(b)], in which each pixel of the map contains a CL spectrum recorded by a CCD camera. The CL map shows intense emission from those bright spots and very little luminescence was observed away from the selected bright spots. The CL spectra extracted from three of the bright spots are shown in Fig. 2(c – e). Each spectrum exhibits a number of resolution limited sharp emission peaks, indicating that excitons are strongly confined in QDs. It appears that the sharp peaks are superimposed on a broad background emission presumably originating from the remaining InGaN quantum well (QW). The narrowest observed peak has a full width at half maximum (FWHM) of ~2 meV, which is limited by the spectral resolution of the CL system (~1.8 meV). Each bright spot in the CL map appears to be associated with multiple sharp emission peaks, possibly indicating the presence of more than one QD.

Figure 3 shows a low-temperature µPL spectrum taken from an unmasked QD sample employing a spot size of ~1 µm. We observed a number of sharp emission peaks centred around 475 nm, confirming the formation of non-polar *a*-plane InGaN QDs, as these sharp peaks appear to be characteristic of the delta-function like density of states in QDs. The narrowest FWHM of these sharp peaks is 1 meV, which is limited by the spectral resolution of the monochromator. Based on the observation of dot-like features in both the CL and PL data, we suggest that during the following growth of a GaN capping layer, re-nitridation of the metallic indium/gallium droplets with ammonia (possibly via interdiffusion with the GaN capping layer) results in the formation of non-polar *a*-plane InGaN QDs within a GaN matrix. This is the same mechanism in the formation that was proposed for dot formation employing an $N_2$ anneal on the *c*-plane[14].

In order to study the effect of the internal electric fields on the electron and hole overlap, a time-correlated single photon counting measurement was performed at 4.2 K to estimate the lifetime of excitons in *c*-plane and *a*-plane InGaN QDs grown by the modified droplet epitaxy approach. Typically, *c*-plane QDs emitting at longer wavelengths show very long lifetimes (in excess of 100 ns for dots emitting at 505 nm[15]) and even at shorter wavelengths lifetimes typically exceed a nanosecond[16,17].

For *a*-plane QDs, a typical PL intensity decay spectrum of emission at 481 nm in Fig. 4 shows a mono-exponential decay and a lifetime of 538 ps can be extracted, significantly shorter than we have ever measured at any wavelength in the *c*-plane case. This is a strong evidence of a significantly increased spatial overlap of electron and hole wavefunctions and reduction of the internal electric fields in *a*-plane InGaN QDs as predicted[18]. Despite the observation of an InGaN QW-like background emission in the CL and µPL data, we see little impact of the InGaN QW layer on the decay curve of *a*-plane QDs, unlike in the *c*-plane case[17]. Further µPL analysis will be published elsewhere[19].

In summary, we have demonstrated the growth of non-polar *a*-plane InGaN QDs by MOVPE using a modified droplet epitaxy method. Indium/gallium nanostructures were formed after the annealing of the InGaN epilayer in $N_2$ and re-reacted with ammonia to form QDs upon capping, in a similar process to that which has been observed on the *c*-plane. It appears that the variations in surface morphology and defect density of the underlying non-polar ELOG GaN pseudo-substrate does not have a significant effect on the QD formation. The observation of resolution limited luminescence peaks with FWHM as small as 2 meV at 8 K in CL and 1 meV at 4.2 K in µPL has confirmed the formation and inclusion of non-polar *a*-plane InGaN QDs within a GaN matrix. Much faster lifetimes have been observed in our *a*-plane QDs than in equivalent *c*-plane QDs due to the internal electric fields being significantly reduced. Such non-polar *a*-plane InGaN QDs would exhibit reduced time-jitter of emission in single photon source applications.

# ACKNOWLEDGMENTS

This work has been funded by the EPSRC (Grant No. EP/H047816/1 and EP/J001627/1).

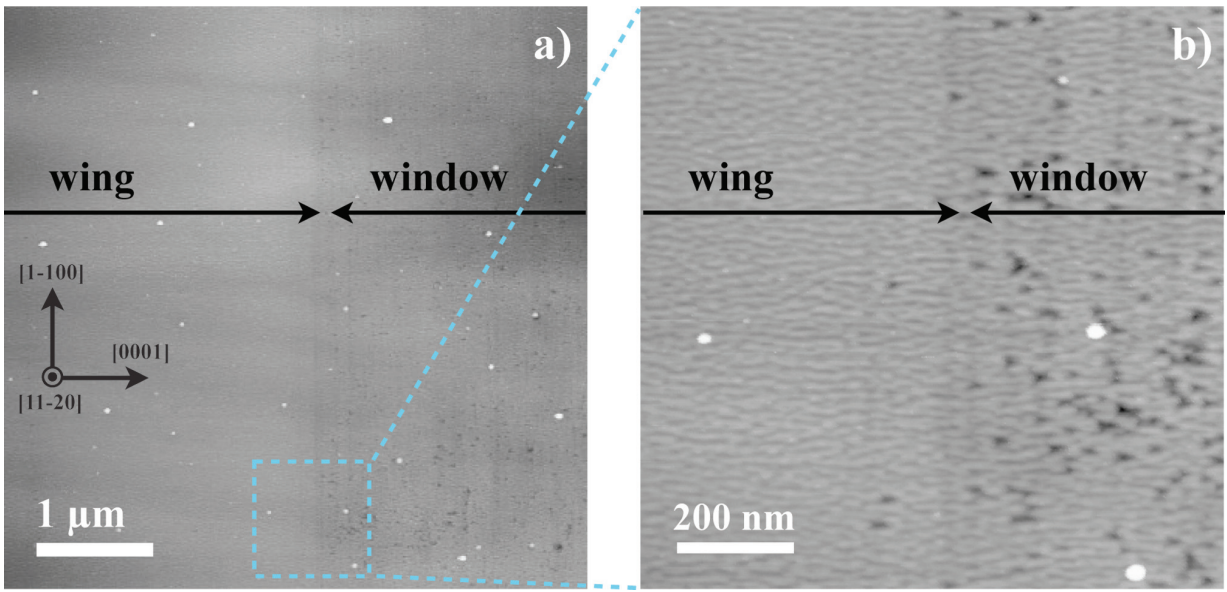

**Fig. 1.** AFM images of a) an overview of InGaN epilayer grown on *a*-plane ELOG GaN following a 30 s anneal in N$_2$ at 695 $^\circ$C (**h**=25 nm), b) a more magnified view over the +*c*-wing and window interface region showing small nanostructures formed on the surface and a network of interlinking InGaN stripes aligned along the [0001] direction (**h**=15 nm).

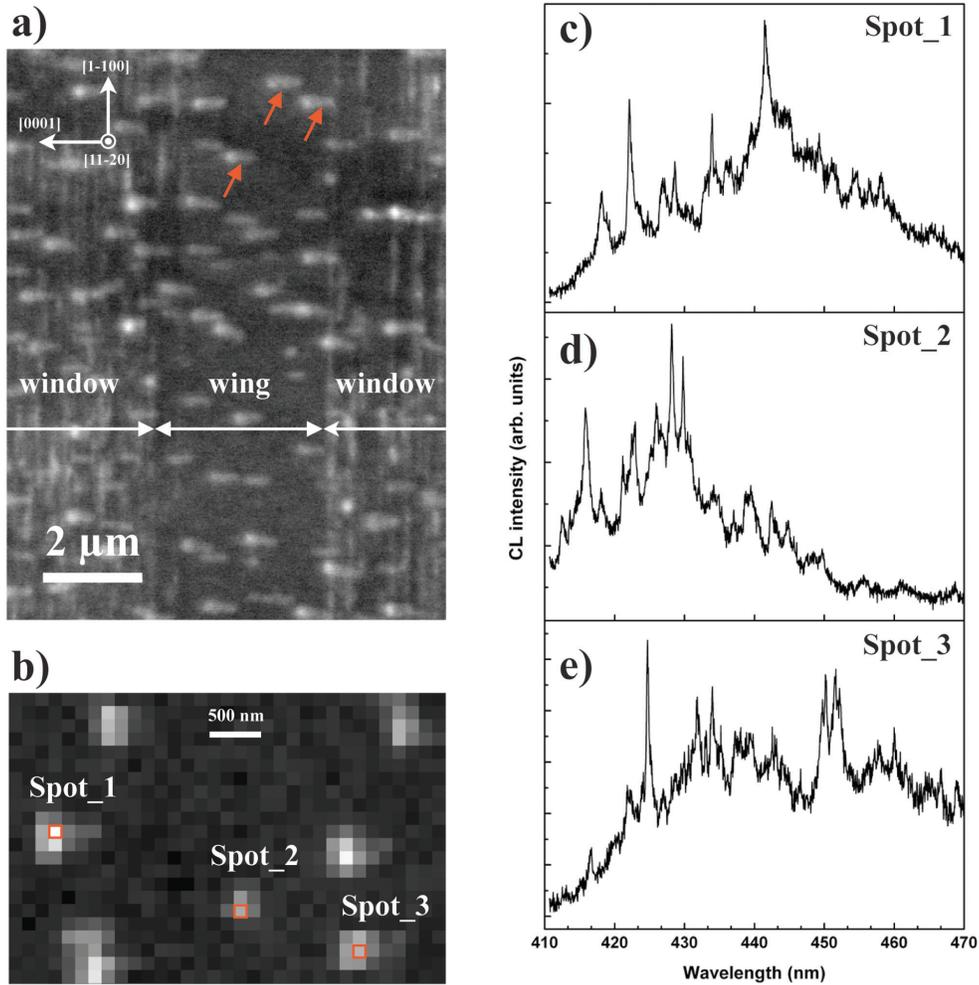

**Fig. 2.** CL images and spectra taken from capped *a*-plane InGaN QDs at 8 K: a) panchromatic CL image, b) CL spectrum image (a two-dimensional CL intensity map of a three-dimensional data set) of 33 x 24 pixels (180 nm) and 2 s of integration time showing spatially localized bright spots corresponding to QDs, c – e) CL spectra extracted from bright spot 1, 2, and 3, respectively, revealing spectral resolution limited sharp peaks.

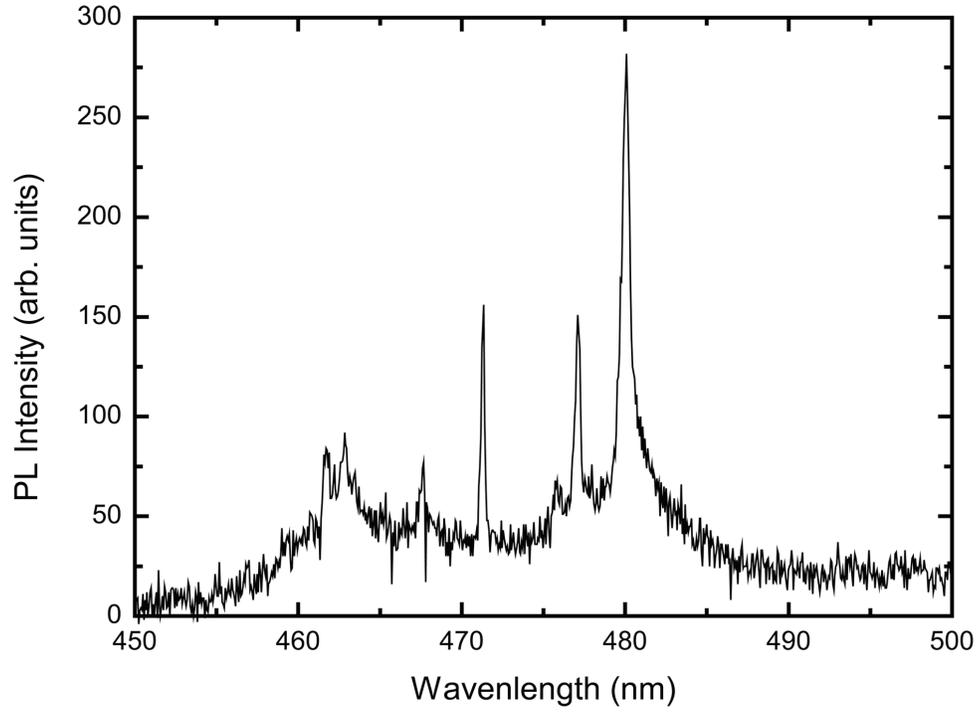

**Fig. 3.** A μPL spectrum from masked *a*-plane InGaN QD sample with a 1 μm spot size and an excitation power density of 7.64 MWcm$^{-2}$ using two-photon excitation at 4.2 K.

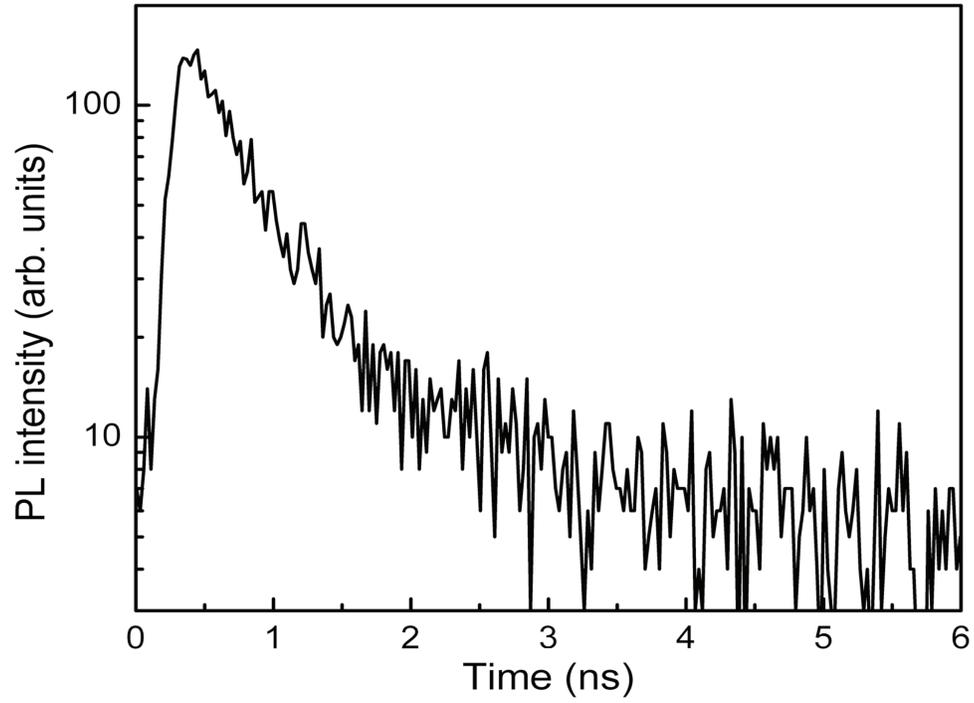

**Fig. 4.** PL intensity decay spectrum recorded at 4.2 K for a non-polar *a*-plane InGaN QD emitting at 481 nm.